 \documentclass[10pt,preprint]{aastex}  

\def\ang{\AA}
\def\arcsec{\hbox{$^{\prime\prime}$}}

\def\gapprox{\lower.4ex\hbox{$\;\buildrel >\over{\scriptstyle\sim}\;$}}
\def\lapprox{\lower.4ex\hbox{$\;\buildrel <\over{\scriptstyle\sim}\;$}}

\def\ref#1{\par\noindent\hangindent1cm {#1}}

\shortauthors{ASCHWANDEN AND BOERNER 2011}
\shorttitle{SOC}

\begin{document}

\title{		Solar Corona Loop Studies with AIA:
		I. Cross-Sectional Temperature Structure	}

\author{        Markus J. Aschwanden and Paul Boerner }

\affil{         Lockheed Martin Advanced Technology Center,
                Solar \& Astrophysics Laboratory,
                Org. ADBS, Bldg.252,
                3251 Hanover St.,
                Palo Alto, CA 94304, USA;
                e-mail: aschwanden@lmsal.com}

\begin{abstract}
We present a first systematic study on the cross-sectional temperature 
structure of coronal loops using the six coronal temperature filters of
the Atmospheric Imaging Assembly (AIA) instrument on the Solar
Dynamics Observatory (SDO). We analyze a sample of 100 loop
snapshots measured at 10 different locations and 10 different
times in active region NOAA 11089 on 2010 July 24, 21:00-22:00 UT.
The cross-sectional flux profiles are measured and a cospatial
background is subtracted in 6 filters in a temperature range of
$T \approx 0.5-16$ MK, and 4 different parameterizations
of differential emission measure (DEM) distributions are fitted.
We find that the reconstructed DEMs consist predominantly of
narrowband peak temperature components with a thermal width of 
$\sigma_{log(T)} \le 0.11\pm0.02$, close to the temperature resolution 
limit of the instrument, consistent with earlier triple-filter analysis
from TRACE by Aschwanden and Nightingale (2005) and from EIS/Hinode 
by Warren et al.~(2008) or Tripathi et al.~(2009). 
We find that 66\% of the loops could be
fitted with a narrowband single-Gaussian DEM model, and 19\% with a 
DEM consisting of two narrowband Gaussians (which mostly result from
pairs of intersecting loops along the same line-of-sight).
The mostly isothermal loop DEMs allow us also to derive an
improved empirical response function of the AIA 94 \ang\ filter,
which needs to be boosted by a factor of $q_{94} =  6.7\pm 1.7$
for temperatures at $log(T) \lapprox 6.3$. 
The main result of near-isothermal loop cross-sections is not
consistent with the predictions of standard nanoflare scenarios,
but can be explained by flare-like heating mechanisms that drive
chromospheric evaporation and upflows of heated plasma coherently
over loop cross-sections of $w \approx 2-4$ Mm.
\end{abstract}


\section{       INTRODUCTION 			}

The plasma dynamics in the solar corona is controlled by the magnetic
field, which channels plasma flows along one-dimensional (1-D) fluxtubes
due to the low plasma-$\beta$ parameter that prevails in most parts of the
corona. This basic 1-D transport process organizes the solar corona into 
bundles of fluxtubes along open or closed magnetic field lines, which we 
generically call ``coronal loops'' (for an overview, e.g., see Aschwanden 2006). 
The fine structure and composition 
of coronal loops, however, is still a subject of debate, which has culminated
into two schools of thought. One theory, mostly inspired by Eugene
Parker (1988), assumes a highly inhomogeneous loop fine structure consisting of
twisted and braided strands that may be produced as a result of microscopic 
magnetic reconnection events, called nanoflares (e.g., see review by 
Klimchuk 2006), which predicts a broad temperature distribution of 
unresolved strands and constitutes a multi-thermal loop. 
Alternatively, loops are thought to have a homogeneous and near-isothermal 
cross-section on a resolved spatial scale of $\gapprox 1$ Mm, if they are 
filled up by a macroscopic chromospheric evaporation process such as the
one known to operate during 
flares (e.g., review by Antonucci et al. 1999). This theory predicts a narrow 
temperature distribution, which is referred to as a monolithic or isothermal 
loop. The determination of the cross-sectional temperature structure,
the prime focus of this paper, is therefore a crucial method to distinguish 
between these two opposite scenarios of microscopic or macroscopic 
plasma heating processes in the solar corona.

The cross-sectional temperature structure of coronal loops 
has been studied early on from 
EUV and soft X-ray images with Skylab, Yohkoh, SMM, SoHO/EIT, and CDS, but 
the spatial resolution of these instruments was limited to a range
of $\approx 2.5\arcsec-10\arcsec$ ($\approx 2-7$ Mm). This is a typical
spatial scale of multi-thermal loop bundles that consist of ensembles
of many unresolved loop strands, which are resolved when inspected with 
high-resolution images, 
such as with TRACE with a pixel size of $0.5\arcsec$ (corresponding to an
effective spatial resolution of $\approx 1.25\arcsec$, i.e., $\approx 0.9$ Mm;
Gburek et al.~2006). Analysis of 
high-resolution TRACE images with three temperature filters in the range of 
$T \approx 0.7-2.7$ MK has given support for near-isothermal loops 
(Del Zanna and Mason 2003; Aschwanden and Nightingale 2005; Warren
et al.~2008, or Tripathi et al.~2009, using also Hinode/EIS data), 
but multi-thermality in loops 
have also been claimed (Schmelz et al.~2009). Each applied method has 
been criticized for different reasons: (i) Loop-associated fluxes can be
heavily contaminated by the multi-thermal background of other loops
along a line-of-sight if the background is not measured cospatially to
the target loop, which unavoidibly leads to a multi-thermal bias;
(ii) triple-filter analysis has a limited temperature
range and thus may not reveal the full temperature width of a differential
emission measure (DEM) distribution, leading to an isothermal bias; 
or (iii) the inversion of a DEM from triple-filter data is underconstrained 
and biased towards the temperature range with the highest instrumental 
sensitivity. All three problems can now be significantly mitigated with 
data from the new {\sl Atmospheric Imaging Assembly (AIA)}  
(Lemen et al.~2011; Boerner et al.~2011) onboard the 
{\sl Solar Dynamics Observatory (SDO)}, which observes the Sun with 8 
different temperature filters, with an uninterrupted cadence of 12 s, 
and a pixel size of $0.6\arcsec$ (corresponding to a spatial resolution of 
$\approx 1.6\arcsec$ or $\approx 1.2$ Mm; Boerner et al.~2001).

In this Paper we present a multi-temperature analysis of 100 loop segments
observed at 10 different spatial locations and 10 different times.
Section 2 contains the description of the data analysis and results
in terms of differential emission measure (DEM) distribution modeling,
while Section 3 contains a discussion of the results and theoretical
consequences, followed by conclusions in Section 4.

\section{       DATA ANALYSIS  			}

\subsection{	Instrument 			}

AIA saw first light on 2010 March 29 and produced since then continuous data
of the full Sun with a $4096 \times 4096$ detector with a pixel size
of $0.6\arcsec$, corresponding to an effective spatial resolution of
$\approx 1.6\arcsec$. AIA contains 10 different wavelength channels,
three in white light and UV, and seven EUV channels, whereof six are
centered on strong iron lines, covering
the coronal range from $T\approx 0.6$ MK to $\gapprox 16$ MK.
AIA records a set of 8 near-simultaneous images in each filter
every 12 s. The number of temperature channels was chosen to be compatible
with the achievable temperature resolution, which is approximately
a Gaussian half width of $\sigma_{\Delta log(T_e)} \approx 0.1$
(corresponding to a full width of half peak of $\Delta log(T_e) \approx 0.25$). 
The lines were chosen to be emitted
by ions of a single element, i.e., iron, to avoid a dependence on the
relative abundances in the coronal plasma. A list of the AIA temperature
channels is given in Table 1. The contributions of different coronal
regions (coronal holes, quiet Sun, active regions, flare plasma) to
the different AIA EUV channels was studied in Boerner et al.~(2001)
and O'Dwyer et al.~(2010),
predicting count rates of $10^1-10^5$ DN s$^{-1}$ for the 6 coronal
AIA channels. The temperature resolution is fundamentally limited due 
to systematic errors in atomic excitation calculations and data noise 
(Judge 2010).

\subsection{	Observations and Loop Selection				}

For the analysis of individual coronal loops we prefer an active
region near disk center, because the line-of-sight or column depth
is smallest at disk center, and thus contains the least confusion
by secondary loops. A suitable active region is NOAA AR 11089, which
crossed the central meridian on 2010 July 24, which we have chosen
for analysis here. In each of the 6 coronal filters we analyze 10 loops
at 10 different times (i.e., 100 events), in time steps of approximately 
6 min, equally spaced during the time interval of 21:00-22:00 UT.

The 100 loop events have 10 different positions, with locations spread 
all over the active region AR 11089. We selected loops with a good 
signal-to-noise ratio, which are mapped out in Fig.~1. The partial image 
shown in Fig.~1 (and all other analyzed images) have a center position 
located at $(-91\arcsec, -413\arcsec)$ from Sun center and a size of 
$(720, 800)$ pixels, corresponding to a field-of-view of 
$(432\arcsec, 480\arcsec)$. All positional information of the displayed 
images (Figs.~1 and 2) and loop coordinates are given in units of the
pixel values $(0<x<720, 0<y<800)$ of the partial images, which can
be downloaded in this format from the {\sl Solar Software (SSW) service 
request} archive of AIA. 

The selection contains 6 cases of loops detected in 6 different
filters at coronal temperatures labeled (after the wavelength channel
in which they were identified) as
131a, 171a, 193a, 211a, 335a 94a) and 4 additional cases chosen
in the loop-rich 171 \ang\ filter (labeled as 171b, 171c, 171d, 171e).
The last case (171e) was deliberately chosen in the same loop structure
as the first case (171a), but in a different segment and distance 
from the loop footpoint. The center coordinates of the selected loop
segments are listed in Table 2. 
The spines of the selected loop segments together
with the encompassing boxes that define the location of the background 
are all shown on the same 171 \ang\ image in Fig.~1, rendered with an 
inverse greyscale (Fig.~1 top) and with a highpass filter 
(Fig.~1, bottom). The loop segments have arbitrary lengths in the range 
of $n_s \approx 20-90$ pixels ($\approx 10-40$ Mm). Curved data stripes 
(aligned with the loop spines) with a width of $n_w=20$ pixels (9 Mm) 
have been extracted for measurements of the loop width $w$, cross-sectional 
flux profiles $F_{\lambda}(x)$, and background profiles $B_{\lambda}(x)$.
Enlarged subimages (with a size of $100 \times 100$ pixels, or 44 Mm$^2$) 
of the selected loops and extracted data stripes are shown in Fig.~2,
along with highpass-filtered renderings to enhance the loop structures.
The enlarged maps show the context of the target loops and neighbored
loop structures. Thus, a total of 
$n = n_L \times n_{\lambda} \times n_t = 600$ subimages were extracted, 
for each of the selected loops 
($n_L=10$), for different times ($n_t=10$), and all six coronal filters 
($n_\lambda=6$).

\subsection{	Solar Rotation 			}

Since we analyze each loop at 10 different times during a time interval
of an hour and the SDO spacecraft has a fixed pointing towards Sun center, 
we have to take the solar rotation into account, which shifts
the loop positions in east-west direction (at the central meridian) 
by an amount of
\begin{equation}
	\Delta x(\Delta t) = {\Delta t \over T_{syn}} 
		{2 \pi (R_{\odot} + h) \over (0.6 \arcsec \times 0.725\ 
		{\rm Mm/\arcsec})} \cos{(b-b_0)} \ \cos{(l-l_0)} \ ,
\end{equation}
with the solar radius $R_{\odot}=696$ Mm, the average altitude of EUV
emission $h \lapprox 0.1 R_{\odot}$, 
the synodic solar rotation period
of $T_{syn}=27.2753$ days, the solar latitude of $(b-b_0) \approx 45^\circ$
and longitude $(l-l_0) \approx 0^\circ$ with respect to Sun center $(l_0, b_0)$, 
amounting up to $\Delta x(\Delta t=1$ hour) $\approx 12$ pixels. 
The corresponding pixel shifts have to be added to the loop coordinates given 
in Table 2, with respect to the reference time of 2010 July 24, 21:00 UT.

\subsection{	Cross-sectional Loop Flux and Background Flux Profiles	}

For each selected loop we mark the start, midpoint, and end position of the
loop spine and interpolate a 2D-spline function with steps of one pixel in order 
to obtain a smooth curvature. A curved array aligned with the positions 
$s_j, j=1,...,n_s$ along the loop spine is then computed in perpendicular
direction $x_i=1,..,n_w$ at each spine point, yielding the flux values
$F_{\lambda}(x_i, s_j)$ in a cartesian grid that corresponds to the stretched-out
loop. The total flux of these stretched-out subimages are shown in Fig.~3 in each
of the 6 wavelength filters for the first time interval $t_1\approx$21:00 UT of
each different loop position. The cross-sectional total flux profiles
are then computed by averaging the 2-D fluxes $F_{\lambda}(x_i, x_j)$ along the
loop spine,
\begin{equation}
	F_{\lambda}^{tot}(x_i) = {1 \over n_s} \sum_{j=1}^{n_s} 
                                 F_{\lambda}^{tot}(x_i, x_j) \ ,
\end{equation}
where the fluxes are given in units of datanumbers per second (DN s$^{-1}$) after
normalizing the counts (DN) by the exposure time, which is different in each filter.
The so-obtained cross-sectional flux profiles $F_{\lambda}^{tot}(x_i)$ are
shown in Fig.~3 (top), which sometimes reveal substantial slopes of the 
cospatial background.  

We characterize the flux profiles with a Gaussian cross-sectional fit 
$G_{\lambda}(x_i)$ that includes a sloped background profile,
\begin{equation}
	G_{\lambda}(x) = F_0 \exp{\left(-{(x-x_0)^2\over 2 \sigma_w^2}\right)}
			+ B_0 + B_1 (x-x_0) \ .
\end{equation}
This 5-parameter fit ($F_0, x_0, \sigma_w, B_0, B_1)$ is executed for each of
the 600 flux profiles (of which 60 are shown in Fig.~3 top for the first time
interval $t_1$). Secondary loops that occasionally appear at the edge of the 
analyzed windows are eliminated by apodization of the Gaussian fit outside of the
flux minima on both sides of the target loop. This method defines a cospatial 
linear background profile,
\begin{equation}
	B_{\lambda}(x) = B_0 + B_1 (x-x_0) \ ,
\end{equation}
and is supposed to remove the loop-unrelated EUV flux along the same line-of-sight
intersecting with the loop center position. The cospatial definition of the loop
and background is crucial for an accurate determination of the loop-associated
flux, while any background evaluation at arbitrary locations away from the loop 
is spoiled by substantial contamination of loop-unrelated flux along the 
line-of-sight. 

The loop centroid positions $x_0$ were found to vary slightly
(in the order of 1-2 pixels) between different temperature filters, which 
indicates some imperfect knowledge of the pointing offset between different
filters, but they do not affect our
results here, because the loop-related flux $F_0$ is evaluated at the
loop centroid position $x_0$ with the Gaussian fit method. 
The knowledge of the pointing offset between the channels is expected
to improve to $<1$ pixel.

The full width at half maximum (FWHM) of the loops, $w$, is related to the
Gaussian widths $\sigma_w$ by, 
\begin{equation}	
	w = 2 \sqrt{ 2 \log{2} }\ \sigma_w \approx 2.35 \ \sigma_w \ .
\end{equation}

We calculated also the cross-correlation coefficients between the 2-D 
stretched-out images $F_{\lambda}(x_i, s_j)$ of each wavelength with that
of the detection wavelength, which are given in each subimage in Fig.~3.
For instance, the loop 171c has been detected in the wavelength of 171 \ang\ ,
but shows no recognizable Gaussian cross-section in 335 \ang\ (Fig.~3 top).
Thus, the flux in the 335 \ang\ channel 
is expected to be uncorrelated with the loop detected in 171 \ang\ .
The cross-correlation coefficient is indeed very low with $CCC(\lambda_{171},
\lambda_{335})=0.193$ (Fig.3). In the calculation of the cross-correlation
coefficients we compensated relative offsets between the wavelength filters
up to $\le 3$ pixels.

\subsection{	Method of Differential Emission Measure Distribution Fits }

The background-subtracted loop fluxes $F_{0\lambda}$ in each of the 6 coronal
wavelength filters are essentially the relevant quantities that constrain the
underlying differential emission measure (DEM) distribution $dEM(T)/dT$ of a
particular loop cross-section,
\begin{equation}
	F_{0\lambda} = \int {dEM(T) \over dT} R_{\lambda}(T) \ dT 
	             = \sum_k EM(T_k) R_{\lambda}(T_k) \ ,
\end{equation}
where $R_{\lambda}(T)$ is the instrumental temperature response function of
each filter $\lambda$. Since the inversion of the DEM is somewhat under-constrained
with a set of 6 flux measurements, we employ a forward-fitting technique with
parameterized DEM distribution functions. 

We choose 4 different parameterizations of DEM functions, which all consist of
superpositions of a variable number of Gaussians as a function of the logarithmic
temperature, but all have 6 free parameters: (1) a single-Gaussian 
($EM_0, T_0, \sigma_{T0}$) with 3 additional fixed Gaussians with amplitudes
$EM_1, EM_2, EM_3$
at fixed temperatures $log(T_k)=5.7+0.4(k-1)$, $k=1,...,3$; (2) a double-Gaussian 
($EM_0, EM_1, T_0, T_1, \sigma_{T0}, \sigma_{T2}$), 
(3) a six-Gaussian DEM ($EM_k, k=1,...,6$ at 6 pre-defined temperatures 
$\log(T_k)=5.7+0.2 (k-1), k=1,..,.6$ with narrow temperature widths $\sigma_T=0.1$),
and (4) the same six-Gaussian DEM function with broad temperature widths
$\sigma_T=0.5$: 
\begin{equation}
	EM(T) = \sum_{k=1}^n EM_k 
		\exp{\left(- {[\log(T)-\log(T_k)]^2 
		\over 2 \sigma_{k}^2}\right)}
		\left\{
	\begin{array}{ll}
	& n=4,\ {\rm single-Gaussian\ (EM_1, T_1, \sigma_1, EM_2, EM_3, EM_4)} \\
	& n=2,\ {\rm double-Gaussian\ (EM_1, T_1, \sigma_1, EM_2, T_2, \sigma_2)}\\
	& n=6,\ {\rm narrow\ six-Gaussian \ (EM_1,...,EM_6, \sigma_T=0.1)} \\
	& n=6,\ {\rm broad\ six-Gaussian \ (EM_1,...,EM_6, \sigma_T=0.5)}
	\end{array}
	\right.
\end{equation}
Examples of the four DEM parameterizations are shown in Fig.~4, where a fit
of each parameterization to an observed DEM (Brosius et al.~1996) is visualized.
The choice of these 4 parameterizations is motivated by the aim to cover the whole
range from isothermal to multi-thermal distributions, but with the same number 
of six free parameters matching the six flux constraints.
The first 3 models with 1-6 Gaussians can all mimic near-isothermal DEM distributions,
while all 4 models can also represent broadband multi-thermal DEM distributions. 
Because the average temperature steps between the peak response of the different
filters is $\Delta \log(T) \approx 0.2$, we have to limit the Gaussian width in
the DEM models to $\sigma_T \gapprox 0.1$. In addition, in order to avoid 
under-constrained solutions outside the sensitivity range of the 6 coronal
AIA temperature filters we set also limits of $5.7 \le \log(T) \le 7.0$.

Our forward-fitting procedure uses a model $F_\lambda^{mod}(x)$
than consists of a convolution of the spatial Gaussian loop cross-sectional profile 
$g(x)=[G_\lambda(x)-B_{\lambda}(x)]/F_0$ with the temperature DEM functions $EM(T)$ 
\begin{equation}
	F^{mod}(x, \lambda) 
	= \exp{\left(-{(x-x_{0,\lambda})^2\over 2 \sigma^2_{w,\lambda}}
			\right)} \ \sum_k EM(T_k) R_\lambda(T_k) \ ,
\end{equation}
which is then fitted to the observed cross-section profiles $F^{obs}(x, \lambda)$
(Eq.~2), using the $\chi^2$-criterion,
\begin{equation}
	\chi^2_{red} = {1 \over (n-n_{free})} \sum_{i,j} 
	{(F^{obs}(x_i, \lambda_j) - F^{mod}(x_i, \lambda_j)^2
	\over \sigma^2(\lambda_j)}  \ ,
\end{equation}
where $n=n_w n_\lambda = 20 \times 6$ is the number of measurements,
$n_{free} = 6$ is the number of free variables of the model, and
the expectation value of the uncertainty $\sigma(\lambda_j)$ in each wavelength 
filter is estimated from the mean standard deviations of the Gaussian cross-section
profile fits $G_{\lambda}(x)$ (Eq.~3) to the observed values $F_{\lambda}(x)$
(Eq.~2) without DEM constraints,
\begin{equation}
	\sigma(\lambda_j)=\sqrt{ {1 \over n_w} 
	\sum_{i} \left[G_{\lambda_j}(x_i) - F_{\lambda_j}^{tot}(x_i)\right]^2} \ .
\end{equation}
This $\chi^2$-criterion yields a value of exactly unity if a perfect solution
of the DEM is found that matches each loop flux. In reality we expect some
uncertainties in the background subtraction that result into inconsistent
flux values that cannot perfectly be fitted by any DEM. Since the expectation values 
of the uncertainties given in Eq.~(10) are empirically determined, it includes all 
possible deviations from a perfect fit of a Gaussian spatial cross-section, 
which includes the Poisson noise of photons, small fluctuations due to real loop 
substructures (e.g., loop strands), or any instrumental effects. 

We encoded our forward-fitting
procedure using the {\sl Interactive Data Language (IDL)} routine {\sl POWELL},
which is based on Powell's method of minimization of functions in multi-dimensions
(Press et al.~1986). The convergence of the Powell method in multiple dimensions
often depends on the accuracy of the initial guess values. For the initial guesses
we estimate first the temperature and emission measure of the DEM peak by an 
inversion of simplified response functions, approximated by a single peak,
which yields the approximate diagonal values of the inversion matrix. 
We test the convergence behavior of our code also by optimization of the linear
as well as logarithmic emission measure and find identical convergence values. 
A test series of the reliability of our DEM inversion method is described
in Appendix A. 

\section{			RESULTS 			}

\subsection{ 			DEM Fits			}

In total we fitted 100 loop events with 4 different DEM models, each one 
simultaneously in 6 filters and 20 spatial positions across the loop 
cross-section. 
We show the best-fit solutions of these 100 loop events at the loop centroid
position $x=x_0$ with the 2-Gaussian DEM model in Fig.~5. The total flux is
generally matched in each filter within a few percents accuracy, although most
loops show a significant evolution during the analyzed time interval.

We express the goodness-of-fit with the $\chi^2$-criterion as defined in
Eq.~(9) and sample the distributions of their values in form of histograms
in Fig.~6, for each of the 4 DEM models separately. All methods are bound
by a lower value of $\chi^2 \gapprox 1$ for the best fits, which is expected
according to our definition of expectation values of uncertainties.

If we choose $\chi^2 \le 2$ as a criterion for acceptable fits,
the single-Gaussian DEM model yields 66\% acceptable fits, 
the double-Gaussians DEM model yields 85\% acceptable fits, 
the narrowband six-Gaussian DEM yields 50\%, and the broadband 
six-Gaussian DEM model only 20\%, while 14\% of the cases have no
acceptable fits with any of the four models. This means that
66\% of the cases can be fitted with a near-isothermal DEM model,
19\% are best fitted with two isothermal components, most likely
representing cases with two intersecting loops of different temperatures
along the same line-of-sight (as verified in one case). Thus 85\% of
the cases are consistent with a single or a pair of isothermal loops.
Only 15\% of the cases could not be fitted with isothermal DEM models,
but do not fit broadband Gaussian DEM models or six-point spline DEM
models neither in 14\% of all cases. This means that 14\% need either
a different DEM parameterization or have no DEM solution at all,
because of errors in the total flux, background flux, or calibration
of the filter response function. 

The fact that the double-Gaussian DEM model performs better than the
single-Gaussian or narrowband six-Gaussian DEM model is because the
double-Gaussian DEM model has more 
flexibility to adjust to the main DEM temperature peak. 
The broadband six-Gaussian DEM model obviously does not
describe most of the observed DEMs because they seem to be mostly dominated
by narrowband temperature peaks. The statistics of the $\chi^2$-values of the
100 fits with 4 different DEM models is also listed in Table 3, showing the
range of goodness-of-fit obtained in each time interval for the 10 loop
locations. The double-Gaussian DEM model has the best statistics with
$\chi^2=1.6\pm0.6$, while the single-Gaussian ($\chi^2=2.0\pm0.7$)
and narrow six-Gaussian DEM Model ($\chi^2=2.3\pm1.0$) have an intermediate
fit quality, and the broadband six-Gaussian DEM model ($\chi^2=3.8\pm1.7$)
has the worst fit quality. 

All 100 DEM fits of the double-Gaussian DEM model are shown in Fig.~7, 
grouped for each spatial location separately. Each panel in Fig.~7
shows the double-Gaussian DEM fit for 10 different times during one hour
of observations. The time evolution of most loops appears to be gradual
without changing the characteristics of the DEM model (except perhaps
for events 211a and 335a). Most loops have a stable narrowband temperature 
structure of a primary loop, which occurs alone (events 131a, 171a, 94a, 171d), 
sometimes in presence of a weaker secondary loop with a different temperature
(events 193a, 211a, 335a, 171b, 171c, 171e). 
None of the 100 cases exhibits a broadband temperature distribution, although
the double-Gaussian DEM model can represent it with the free parameters of
variable temperature widths ($\sigma_{T1}, \sigma_{T2}$).

In Fig.~8 we show a comparison of different DEM models for all cases with an
acceptable goodness-of-fit, say $\chi^2 \le 1.5$. In 23 cases (out of the 100
events) we find that at least 3 DEM models have an acceptable fit, which are
shown in Fig.~8. The fact that these DEM models have all acceptable fits,
although different parameterizations, implies that they are all consistent
with the data, and thus reveal realistic uncertaintities in the DEM definitions.
Most cases show convergence to very similar single or multi-peaked DEM
distributions. Only in 6 out of the 23 good solutions we find that a broadband
DEM distribution is also consistent with the data, which is only the case if
the other DEM models show a double or triple-peak solution, but never for a
single-peak solution. Thus, there is no evidence for any single-peaked
broadband DEM distribution, while ambiguities between multi-peaked narrowband
and a single-peaked broadband DEM distribution exists only in very few cases.
These results demonstrate a strong statistical trend of narrowband DEM 
distributions for loop cross-sections, similarly as found earlier with
triple-filter analysis from TRACE data (Aschwanden and Nightingale 2005).

\subsection{ 	Physical Parameters 			}

The statistical distributions of the best-fit parameters (of the strongest 
DEM temperature peaks) are shown in Fig.~9 in form of histograms, obtained
from the 2-Gaussian DEM model.
Loop temperatures were found with a mean and standard deviation of
$\log(T_e)=6.1\pm0.4$, which covers the entire range from $T_e=0.4$ MK
to $T_e=5.0$ MK. The overabundance of loop temperatures in the temperature
range of $T_e=0.50-0.63$ MK is a selection effect, because we arbitrarily
choose half of the identified loops in the 171 \ang\ filter. 
The electron density is found in a
range of $\log(n_e)=8.9\pm0.2$, based on the measurements of the 
loop emission measure $EM$, width $w$, and assumption of unity filling factor,
\begin{equation}
	n_e = \sqrt{ {EM \over w} } \ .
\end{equation}
The loop widths were found in a range of $w=2.9\pm0.8$ Mm, which corresponds
to a range of $\approx 5-10$ pixels (Table 2). 

\subsection{	Empirical Response Function for the 94 A filter 	}

The nominal response function of the AIA 94 \ang\ filter as available in the
{\sl Solar Software (SSW)} package is known to be incorrect, according to
information from James Lemen, Harry Warren, Joan Schmelz, Nancy Brickhouse,
and Peter Beiersdorfer (private communication). The nominal 94 \ang\ response
function is shown in Fig.~10, which shows a double peak at $log(T) \approx 6.1$
due to Fe X lines and at $\log(T) \approx 6.8$ due to Fe XVIII lines. 
AIA images in 94 \ang\ often display strong emission from the $\log(T) \approx 6.0$
quiet corona that is in excess of the expected response function. It is therefore
suspected that a large numer of Fe X atomic transitions are not included in the
currently available CHIANTI code, which is the atomic database of the AIA
response function calculation. We experienced this problem directly in the
sense that the flux observed in the 94 \ang\ filter could not be matched with 
the nominal response function in our statistical DEM modeling of 100 mostly
isothermal loop events.

In order to derive a first-order correction to the nominal 94 \ang\ response
function $R_{94}(T)^{nom}$ we defined an empirical boost factor $q_{94}$ for 
the cool-temperature peak of the 94 \ang\ response function,
say at $\log(T) \le 6.3$,
\begin{equation}
	R_{94}(T)^{emp} = \left\{
		\begin{array}{ll}
		q_{94} R_{94}(T)^{nom} & {\rm for}\ \log(T) \le 6.3 \ {\rm \ang} \\
		\quad  R_{94}(T)^{nom} & {\rm for}\ \log(T) > 6.3 \ {\rm \ang} 
		\end{array}
		\right.
\end{equation}
In order to determine the correction factor $q_{94}$ we ran first all DEM fits
for the selected 100 loop events with the free variable $q_{94}$ in addition to 
the list of six free DEM parameters defined with Eqs.~(7) and (8) and obtained
a statistical distribution of best-fit values $q_{94}$ with a mean and standard
deviation of,
\begin{equation}
	q_{94} = 6.7 \pm 1.7 \ ,
\end{equation}
which is shown in Fig.~10 (left). This correction factor is most accurately
obtained from near-isothermal cases with temperatures around 
$\log(T) \approx 6.0$. We fixed then this correction factor to a
constant value $q_{94}=6.7$ and repeated all the fits with the corrected
empirical response function $R_{94}(T)^{emp}$ specified in Eq.~(12), leading
to the results described in Section 3 and Figures 5-9. 

There is also some concern about the accuracy of the 131 \ang\ response 
function, 
where the ionization fractions for Fe VIII have been called into question. 
Young et al.~(2007) have noted that Si VII and Fe VIII spectroheliograms 
look nearly identical despite the fact that these two ions are separated 
in temperature by 0.2 dex. However, we fitted our DEMs with and without
the 131 \ang\ channel without finding a significant difference. 

\section{       DISCUSSION 					}

We discuss out results in view of instrumental biases, such as
the background subtraction issue (Section 4.1), the narrowband
temperature filter bias (Section 4.2), and put the results in
context of physical loop models (Section 4.3).

\subsection{	The Loop and Background Co-spatiality Issue 		}

The thermal structure of an individual coronal loop can only be 
properly determined if the loop-associated EUV flux 
$F_{\lambda}^{loop}=F_{\lambda}^{tot}-B_{\lambda}$ 
is accurately separated from the background flux $B_{\lambda}$ 
along each line-of-sight. The background corona is multi-thermal,
since it consists of tens to thousands of other loops along each 
line-of-sight, any contamination of loop fluxes $F_{\lambda}^{loop}$
with background fluxes $B_{\lambda}$ will add also a multi-thermal
contribution to the resulting DEM. Quantitative modeling of the
multi-loop background corona that reproduces the observed DEMs has
been performed with the {\sl CELTIC} model (Aschwanden et al.~2007). 
The best background 
subtraction is achieved when the background is evaluated cospatially 
to the loop, say with a linear interpolation between the left 
and right-hand side of a loop cross-section, which is essentially the
technique we applied in our study here. Other methods that estimate 
the background
flux not cospatially to the loop are fundamentally flawed, because 
the coronal flux at two different locations is unrelated to each other.
It is therefore imperative to provide the location and definition
of background fluxes in loop studies. 

A recent study analyzed a coronal loop with AIA, observed on 2010 Aug 3,
and carried out some DEM modeling, claiming a multithermal rather than
an isothermal DEM distribution (Schmelz et al.~2010). Unfortunately
this study does not specify the loop coordinates, total fluxes, 
background fluxes, or background locations. The description that
{\sl 10 pixels from a clean background area} were subtracted from
the loop fluxes suggests a non-cospatial determination of loop and
background fluxes, which most likely contributes a multi-thermal 
contamination to the loop and this way explains the inferred result 
of a multithermal loop. The cospatial total fluxes $F_{\lambda}^{tot}$
and background fluxes $B_{\lambda}$ shown for 100 loop events in
Fig.~5 demonstrate that fluxes from the most obvious (foreground) loops 
amount only to about $5\%-50\%$ of
the total flux in each pixel, and thus a slight contamination
changes the flux ratios in different filters and the resulting DEMs
significantly. A fit of a broadband DEM with a temperature
width of $\Delta log(T) \approx 0.6$ is shown in Fig.~4 of Schmelz
et al.~(2010), which obviously fits better than a narrowband DEM. However,
the inferred DEM applies not to the loop alone, but rather to
a combination of the target loop plus an unknown fraction of the background
corona. Our results of dominantly narrowband DEMs for 100 loop
events is therefore not in conflict with that study, because we would
obtain similar broadband DEMs with non-cospatial background estimates.

\subsection{	Narrowband and Broadband Temperature Diagnostics      }

The issue of the inadequacy of temperature measurements in the solar corona
through narrowband filter and line ratios was raised in Martens et al~(2002),
where it was argued that recently discovered isothermal loops could be
an artifact of narrowband filter methods. This criticism
indeed needs to be taken seriously, because data with a narrow
temperature coverage may not have sufficient flexibiltiy to reconstruct
broadband-temperature DEM distributions, such as triple-filter data from
TRACE within a combined temperature range of $T\approx 0.7-2.7$ MK. 
Although instruments with a broader temperature range exist (e.g.,
SoHO/CDS or Hinode/EIS), reliable multi-wavelength modeling is extremely
difficult due to insufficient spatial resolution, inconsistent instrument 
calibrations, non-overlapping time coverage, and image distortion of 
scanning spectrographs. A single instrument such as SDO/AIA with a sufficient
large number of temperature filters, self-consistent calibration, 
high spatial resolution, temporal simultaneity, and sufficient cadence
is therefore the first good opportunity to attempt reliable multi-temperature
modeling of coronal loops.

Despite the incompleteness of the 94 \ang\ response function, which we
bootstrapped with self-consistent solutions with the other 5 coronal AIA
filters, AIA essentially allows us to model DEMs in a temperature range
of $log(T) \approx 5.7-7.2$ (or $T=0.5-16$ MK), which spans a logarithmic
range of 1.5 decades. The average temperature spacing is $\Delta log(T)
\approx 1.5/6 = 0.25$ dex, which defines a temperature resolution with
a Gaussian width of $\sigma_T \approx 0.1$ dex. We have chosen 4 different
parameterizations of DEM distributions that can resolve narrowband
temperature peaks down to this resolution of $\sigma_T \gapprox 0.1$, and
at the same time can represent broad distributions up to the full
temperature sensitivity range of $\Delta log(T) \approx 1.5$ dex.
Interestingly, the majority of reconstructed DEMs converged to narrowband
temperature peaks near $\sigma_T \approx 0.1$ dex (Fig.~9), especially
those DEMs with the most reliable goodness-of-fit $\chi^2$-values
($\chi^2=1.6\pm0.6$, see Table 3 for double-Gaussian DEMs).
A similar result was found with TRACE triple-filter DEM modeling,
where 84\% of acceptable DEMs (with $\chi^2 \le 1.5$) were found to be
isothermal (Aschwanden and Nightingale 2005), despite of the smaller
number of available temperature filters.
Could these results be an artifact of the numerical forward-fitting code?
The fact that 3 different DEM parameterizations converge all to the same
narrow-temperature peaks seems to rule out a numerical artifact. The only
DEM parameterization that we did not allow to converge to narrowband
temperature solutions (i.e., the broadband six-Gaussian DEM function
with Gaussian widths of $\sigma_T = 0.5$), converged in almost all cases
to a DEM solution with a much poorer goodness-of-fit $\chi^2$-value
($\chi^2=3.8\pm1.7$, see Table 3 for broadband six-Gaussian DEMs). 
It therefore appears that broad DEM distributions are very atypical for 
single loops after proper (cospatial) background subtraction, unless multiple
loops intersect each other cospatially. This explains the double-peak
DEM distributions we found in half of the analyzed cases (see Fig.~7).
Our AIA results are also consistent with a study using EIS/Hinode
(Warren et al.~2008), where
a set of 20 coronal loops was found to be isothermal within a very narrow
temperature distribution of $\Delta T \lapprox 0.3$ MK at $T \approx
10^{6.1}-10^{6.2}$ K $\approx 1.4$ MK, which corresponds to a 
Gaussian width $\sigma_{log(T)} = \log(T+\sqrt{2} \Delta T)-\log(T)
\approx 0.11$ dex. In comparison, our result shows an upper limit of
$\sigma_{log(T)} \gapprox 0.11\pm0.02$ (Fig.~9). Our results are also
consistent with Hinode/EIS spectroscopic observations that revealed 
near-isothermal upflows in active region loops (Tripathi et al.~2009).

All these results suggest that AIA enables us reliably to reconstruct
DEMs in a temperature range of $\Delta log(T) \lapprox 1.5$
with a resolution of $\Delta log(T) \gapprox 0.25$, if loops can
be properly separated from the cospatial coronal background.
However, limitations exist when the cospatial cross-sectional background profile
significantly deviates from a linear interpolation, or when substantial density 
and temperature gradients along the averaged loop segment exist,
in which case a self-consistent DEM solution may be inhibited. 
In addition, the Poissonian photon noise and calibration errors
contribute to the uncertainty of DEM fits.

\subsection{	Isothermal versus Multi-thermal Loop Models 	}

How do we explain the result of predominant isothermal loops in terms
of physical models? Let us first discuss isothermality in terms of nanoflare
models. Since nanoflares occur on unresolved spatial scales and have no
cross-field transport, every nanoflare model predicts a highly inhomogeneous
density and temperature structure of macroscopic loops (Klimchuk 2006).
The only way to make a nanoflaring loop structure more isothermal is to
synchronize a storm of nanoflares and to streamline them to identical energy 
outputs, so that their release appears to be simultaneous and homogeneous 
across a loop cross-section, which in the continuum limit is macroscopically
indistiguishable from a monolithic loop. Following Occam's razor,
assuming a coherent isothermal upflow in a single flux tube is a weaker 
assumption than synchronized upflows with equal temperatures in a
multi-thread structure. However, a stronger argument that currently 
supports a fragmented energy release, such as nanoflares,
is the observed lifetime of some coronal loops that are more extended than 
expected for an impulsive heating phase with subsequent
cooling (e.g., Warren et al.~2008). However, the time evolution of isolated
loop strands has first to be studied with high spatial resolution and high
time cadence in many temperature filters, such as AIA provides, before 
clear-cut cases can be established. At this point we can only conclude
that the observed isothermality of coronal loops is not consistent with
standard nanoflare scenarios, nor do nanoflare models explain or predict
the isothermal property.

If an isothermal loop cross-section cannot be produced by nanoflares,
what other physical process can account for it? The most natural
mechanism seems to be that of chromospheric evaporation as known in
flares (e.g., Antonucci et al.~1999), where either coronal or 
chromospheric magnetic reconnection processes cause a local heating 
of the chromosphere and drive coherent upflows of heated plasma into 
a coronal loop conduit. Although the details of the
heating cross-section transverse to the magnetic field is not fully
understood, flare observations yield clear evidence that postflare
loops are filled impulsively with heated plasma over a cross-section
of several thousand kilometers. Distributions of hard X-ray footpoint
sources in flare loops have typical FWHM of $\approx 2\arcsec-8\arcsec$,
or $w \approx 1.5-6.0$ Mm, according to RHESSI measurements with a spatial
resolution of $\approx 2\arcsec$ (Dennis and Pernak 2009). 
One of the few flare observations that resolve the blue-shifted upflows
from near-cospatial red-shifted downflows is described in
Czaykowska et al.~(1999), which yields evidence for coherent 
near-isothermal upflows over a cross-section of $\approx 4\arcsec$
($\lapprox 3$ Mm). Similarly, Hinode/EIS spectroscopic observations
showed near-isothermal upflows in active region loops (Tripathi et al.~2009).
The heating process at a cospatial
location lasts in the order of minutes for flare loops, while a flare
can last for hours, with the energy release propagating along and
perpendicular to the neutral line. Of course, the upflows in flare
loops are more or less continuous during the heating phase, and thus
no thermal equilibrium is reached that obeys the theoretically
expected conductive and radiative cooling times, explaining 
the discrepancies between observed loop lifetimes and theoretically
calculated cooling times (Warren et al.~2008). The temperature
of an arbitrary loop cross-section can be nearly constant
during some part of the heating phase due to the continuous upflow
of isothermal plasma, which explains the slow time evolution observed
in coronal loops during time intervals of hours (see Fig.~7). 
It seems natural to suggest a flare-like heating mechanism for
active region loops, although there might be significant differences.
While the coronal height of magnetic reconnection regions is
established in solar flares (e.g., Masuda et al.~1995), 
reconnection processes may happen in the chromosphere and
transition region (Aschwanden et al. 2007; Gudiksen and Nordlund 2005a,b),
causing subsequent upflows of heated plasma into the coronal parts of
active region loops.  
It is too early to speculate about the details of the generic
heating mechanism of active region loops, before we analyzed
comprehensive multi-wavelength observations of coronal loops such
as with AIA and modeled their hydrodynamic evolution self-consistently.

\section{       CONCLUSIONS 					} 

Our study of the cross-sectional temperature structure of coronal loops
using AIA six-filter data leads us to the following conclusions:

\begin{enumerate}
\item{From an sample of 100 loop snapshots measured at 10
different locations and at 10 different times in an active region
near disk center, we measured the cross-sectional flux profiles in
6 coronal temperature filters, subtracted a cospatial linearly
interpolated background, and reconstructed the differential
emission measure (DEM) distributions in the temperature range of
$T \approx 0.5-16$ MK. We found dominantly narrowband peak
temperature components with a thermal width of $\sigma_{log(T)}
\le 0.11\pm0.02$, close to the temperature resolution limit
of the instrument ($\sigma_{\log(T)} = 0.1$). This result, derived
from 6-filter data, is consistent with similar analysis from
TRACE triple-filter data (Aschwanden and Nightingale 2005) and
from EIS/Hinode data (Warren et al.~2008; Tripathi et al.~2009).}

\item{The DEM distribution was modeled with 4 different parameterizations,
including one or two Gaussians with free parameters and six-Gaussian 
functions with narrowband ($\sigma_{log(T)}\ge 0.1$) and
broadband ($\sigma_{log(T)}\ge 0.1$) temperature components. 
Among the subset of DEM solutions with acceptable goodness-of-fit
($\chi^2 < 2$) we find that the observed fluxes can be fitted with
a single-Gaussian DEMs in 66\%, with double-Gaussian DEMs in 19\%, 
while 14\% could not be fitted with any of the four DEM models,
which might have no DEM solution because of errors in the 6 filter
fluxes, backgrounds, or in the calibration. The cases with double-Gaussian
DEMs are most likely attributed to a pair of loops that intersects at 
the same line-of-sight.}

\item{The nominal response function of the AIA 94 \ang\ filter is found
to be inconsistent with the other 5 coronal temperature filters in the
low-temperature part of $log(T) \lapprox 6.3$. From self-consistent fits
of 100 mostly isothermal loop events we establish an empirical response
function where the low-temperature response is boosted by an average 
correction factor of $q_{94} =  6.7\pm 1.7$, which is consistent with
the known deficiency of missing Fe X lines in the Chianti code.}

\item{The main result of near-isothermal loop structure in this
random sample of 100 loop snapshots provides a strong argument for
a coherently operating heating mechanism across the observed macroscopic
loop structures with a width of $w \approx 2-4$ Mm.   
The observed isothermality of coronal loops is not consistent with
standard nanoflare scenarios, nor do nanoflare models explain or predict
the isothermal property. Flare-like heating mechanisms that drive
chromospheric evaporation and upflow of heated plasma into coronal loops
are known to produce near-isothermal loop cross-sections, and thus
may be a also a viable mechanism for heating of active region loops.}
\end{enumerate}

Future loop studies with AIA are anticipated that determine the thermal
loop structure along the loop axis, as well as a function of time, which
will provide unprecedented input for hydrodynamic simulations of loops.
Time-dependent hydrodynamic models will allow us then to deconvolve the 
temporal and spatial heating function, which ultimately will lead to
the identification of coronal heating mechanisms.

\acknowledgements {\sl Acknowledgements:} This work is partially 
supported by NASA under contract NNG04EA00C of the SDO/AIA instrument.

\section*{ Appendix A : Testing the Inversion of DEM Distributions }

In order to test the reliability of our DEM inversion code we perform a
test by simulating
a set of 21 DEM distributions that contain one or two Gaussian peaks,
which are then converted into loop cross-sectional flux profiles 
$F_{\lambda}^{tot}(x)$ of
the 6 AIA filters $\lambda$, with photon noise added, and inverted with
our forward-fitting model with the double-Gaussians DEM model (Eq.~7). 
Assuming that we averaged the flux values of the loop cross-sections
over a loop length segment of $n_s=50$ AIA pixels, the photon noise
is reduced by a factor of $1/\sqrt{n_s} \approx 1/7$ with respect
to the photon noise in a single pixel.
In the simulated DEMs we choose double peaks with peak temperatures
$T_i = 10^{5.6+0.2*i}$ for $i=1,...,6$. There are 6 single-peak cases
and $(6\times 5)/2=15$ double-peak cases, hence a total of 21 cases,
numbered consecutively in Fig.~11 in order of increasing separation
of the temperature peaks. The Gaussian widths of all components is
$\sigma_{log(T)}=0.1$. The emission measures were chosen as
$EM_1=10 (T_i/T_0)^4$ DN s$^{=1}$ for the first peak, and 
$EM_2=EM_1/2$ for the second peak, following the Rosner-Tucker-Vaiana 
scaling law ($EM \propto T^4$), in order to reproduce typical
emission measures. The loop width was chosen to correspond to 
5 AIA pixels, and the fitted cross-sectional profiles contain 
$n_w=20$ AIA pixels in width and $n_s=50$ AIA pixels in length. 

The results of the inversion and the goodness-of-fit $\chi^2_{red}$
are shown in Fig.~12. We started with initial guess values in the
forward-fitting code based on the input flux values and obtained
an acceptable fit of $\chi^2_{red} \lapprox 2.0$ in most cases. 
In those cases where no acceptable fit was found in the first trial, we
iterated the initial guesses of the temperature values randomly
in the range of $T = 1-10$ MK until we obtained an acceptable fit 
with $\chi^2_{red}$, which was achieved for all cases. 
The mean and standard deviation
of the goodness-of-fit values is $\chi^2_{red}=1.33\pm0.25$ for
the 21 cases shown in Fig.~11. Therefore, we conclude that our
forward-fitting code reliably retrieves the DEM distributions
within the uncertainty of the data noise, in theory. Practically,
there are errors in the measurement of the total fluxes, background
fluxes, and calibration that are not part of this simulation.


\section*{REFERENCES} 

\def\ref#1{\par\noindent\hangindent1cm {#1}} 
\def\aap {{\sl Astron. Astrophys.}\ } 
\def\apj {{\sl Astrophys. J.}\ } 
\def\sp  {{\sl Solar Phys.}\ } 
\def\ssr {{\sl Space Science Rev.}\ } 

\ref{Antonucci,E., Alexander,D., Culhane,J.L., DeJager,C., MacNeice,P., 
	Somov,B.V., and Zarro,D.M. 1999, {\sl Chapter 10: Flare dynamics}, 
	in {\sl The Many Faces of the Sun: A Summary of the Results from 
	NASA's Solar Maximum Mission}, (eds. Strong,K.T., Saba,J.L.R., 
	Haisch,B.M., and Schmelz,J.T.) 610p., Springer, Berlin, p.331.}
\ref{Aschwanden, M.J., Winebarger, A., Tsiklauri, D., and Peter, H.
	2007, \apj 659, 1673.}
\ref{Aschwanden, M.J. and Nightingale, R.W. 2005, \apj 633, 499.}
\ref{Aschwanden, M.J. 2006,
	{\sl Physics of the Solar Corona - An Introduction with Problems 
	and Solutions}, Springer/Praxis, New York, 
	ISBN 3-540-30765-6, paperback, 892p.}
\ref{Aschwanden, M.J., Nightingale, R.W., and Boerner, P. 2007, \apj 656, 577.}
\ref{Boerner, P., Edwards, C., Lemen, J., Rausch, A., Schrijver, C.,
	Shine, R., Shing, L., Stern, R., Tarbell, T., Title, A.,
	and Wolfson, C.J. 2011, {\sl Initial calibration of the Atmospheric
	Imaging Assembly Instrument}, (in preparation).}
\ref{Czaykowska, A., De Pontieu, B., Alexander, D., and Rank, G.
	1999, \apj 521, L75.}
\ref{Del Zanna, G. and Mason, H.E. 2003, \aap 406, 1089.}
\ref{Dennis, B.R. and Pernak, L. 2009, \apj 698, 2131.}
\ref{Gburek, S., Sylwester, J., and Martens, P. 2006, \sp 239, 531.}
\ref{Gudiksen, B.V., and Nordlund, A. 2005a, \apj 618, 1020.}
\ref{Gudiksen, B.V., and Nordlund, A. 2005b, \apj 618, 1031.}
\ref{Judge, P.G. 2010, ApJ 708, 1238.}
\ref{Klimchuk, J.A. 2006, \sp 234, 41.}
\ref{Lemen, J. and AIA Team, 2011, \sp (in preparation).}
\ref{Martens, P.C.H., Cirtain, J.W., and Schmelz, J.T. 2002, \apj 577, L115.}
\ref{Masuda,S., Kosugi.T., Hara,H., Sakao,T., Shibata,K., and Tsuneta,S.
 	1995, PASJ 47, 677.}
\ref{O'Dwyer, B.O., Del Zanna, G., Mason, H.E., Weber, M.A., and Tripathi, D.
        2010, A\&A (subm).}
\ref{Parker, E.N. 1988, \apj 330, 474.}
\ref{Press, W.H., Flannery, B.P., Taukolsky, S.A., and Vetterling, W.T.
	1986, {\sl Numerical Recipes - The Art of Scientific Computing},
	Cambridge University Press, Cambridge.}
\ref{Schmelz, J.T., Naraoui, K., Rightmire, L.A., Kimble, J.A., Del Zanna, G.,
	Cirtain, J.W., DeLuca, E.E., and Mason, H.E. 2009, \apj 691, 503.}
\ref{Schmelz, J.T., Kimble, J.A., Jenkins, B.S., Worley, B.T., Anderson, D.J.,
	Pathak, S., and Saar, S.H. 2010, \apj 725, L34.}
\ref{Tripathi, D., Mason, H.E., Dwivedi, B.N., DelZanna, G., and Young, P.R.
	2009, ApJ 694, 1256}
\ref{Warren, H.P., Ugarte-Urra, I., Doschek, G., Brooks, D.H., and
	Williams, D.R. 2008, \apj 686, L131.}
\ref{Young, P.R., DelZanna, G., Mason, H.E., Dere, .P., Landi, E.,
	Landini, M., Doschek, G.A., Brown, C.M., Culhane, L., Harra, L.K.,
	Watanaba, T., and Hisohisa, H. 2007, Publ. Astron. Soc. Japan 
	59, S857.}

\clearpage

\begin{deluxetable}{lll}
\tabletypesize{\normalsize}
\tablecaption{AIA/SDO wavelength bands.}
\tablewidth{0pt}
\tablehead{
\colhead{channel name}&
\colhead{Primary ions}& 
\colhead{Characteristic}\\
\colhead{}&
\colhead{}&
\colhead{log(T)}}
\startdata
white light     &continuum              &3.7            \\
1700 \ang       &continuum              &3.7            \\
304 \ang        &He II                  &4.7            \\
1600 \ang       &C IV+cont.             &5.0            \\
171 \ang        &Fe IX                  &5.8            \\
193 \ang        &Fe XII, XXIV           &6.1, 7.3       \\
211 \ang        &Fe XIV                 &6.3            \\
335 \ang        &Fe XVI                 &6.4            \\
94 \ang         &Fe XVIII               &6.8            \\
131 \ang        &Fe VIII, XXI           &5.6, 7.1       \\
\enddata
\end{deluxetable}


\begin{deluxetable}{ccccc}
\tabletypesize{\normalsize}
\tablecaption{Center coordinates (at reference time 2010 Jul 24, 21:00 UT), 
lengths, and widths of selected loops.}
\tablewidth{0pt}
\tablehead{
\colhead{Loop name}&
\colhead{Position}& 
\colhead{Segment length}&
\colhead{Segment width}&
\colhead{Loop width}\\
\colhead{}&
\colhead{$(x_0,y_0)$}&
\colhead{$(n_s)$}&
\colhead{$(n_w)$}&
\colhead{$w$}\\
\colhead{}&
\colhead{(pixels)}&
\colhead{(pixels)}&
\colhead{(pixels)}&
\colhead{(pixels)}}
\startdata
131a	  &(140,404)	&21		&20		&8.2\\
171a	  &(463,264)	&89		&20		&8.1\\
193a	  &(235,218)	&35		&28		&8.0\\
211a	  &(298,286)	&20		&20		&4.9\\
335a      &(329,296)    &21		&20		&6.9\\
 94a      &(302,417)    &30		&20		&9.7\\
171b      &(150,266)    &33		&20		&7.2\\
171c      &(203,425)    &30		&20		&5.8\\
171d      &(462,235)    &25		&20		&4.7\\
171e      &(112,431)    &34 		&20		&7.5\\
\enddata
\end{deluxetable}


\begin{deluxetable}{ccccc}
\tabletypesize{\normalsize}
\tablecaption{Range of goodness-of-fit $\chi^2$-values obtained
for the time range of 2010 Jul 24, 21:00-22:00 UT for each of the
10 loop positions and 4 different DEM models.} 
\tablewidth{0pt}
\tablehead{
\colhead{Loop name}&
\colhead{1-Gaussian DEM}& 
\colhead{2-Gaussian DEM}&
\colhead{Narrow 6-Gaussian DEM}&
\colhead{Broad 6-Gaussian DEM}}
\startdata
131a &  1.32$-$ 2.40 &  1.09$-$ 1.86 &  1.16$-$ 1.99 &  1.73$-$ 3.77 \\
171a &  1.54$-$ 3.10 &  1.44$-$ 2.86 &  1.75$-$ 3.73 &  2.58$-$ 6.52 \\
193a &  1.63$-$ 2.97 &  1.48$-$ 1.75 &  1.79$-$ 3.33 &  2.37$-$ 5.76 \\
211a &  1.39$-$ 3.57 &  1.14$-$ 3.78 &  1.99$-$ 3.61 &  5.44$-$ 7.76 \\
335a &  1.90$-$ 3.28 &  1.14$-$ 2.79 &  1.94$-$ 4.58 &  3.03$-$ 6.43 \\
 94a &  1.26$-$ 4.09 &  1.13$-$ 3.83 &  1.86$-$ 7.60 &  3.50$-$ 6.36 \\
171b &  1.21$-$ 1.62 &  1.09$-$ 1.39 &  1.09$-$ 1.35 &  1.16$-$ 1.72 \\
171c &  1.08$-$ 1.89 &  1.08$-$ 1.67 &  1.08$-$ 1.91 &  1.41$-$ 4.14 \\
171d &  1.33$-$ 1.98 &  1.16$-$ 1.76 &  1.52$-$ 4.42 &  1.54$-$ 3.78 \\
171e &  1.15$-$ 1.51 &  1.11$-$ 1.46 &  1.17$-$ 1.61 &  2.74$-$ 5.15 \\
     &               &               &                  &               \\
Average & 1.95$\pm$0.68 & 1.59$\pm$0.58 & 2.27$\pm$1.06 & 3.76$\pm$1.70 \\
\enddata
\end{deluxetable}


\begin{figure}
\centerline{\includegraphics[width=\textwidth]{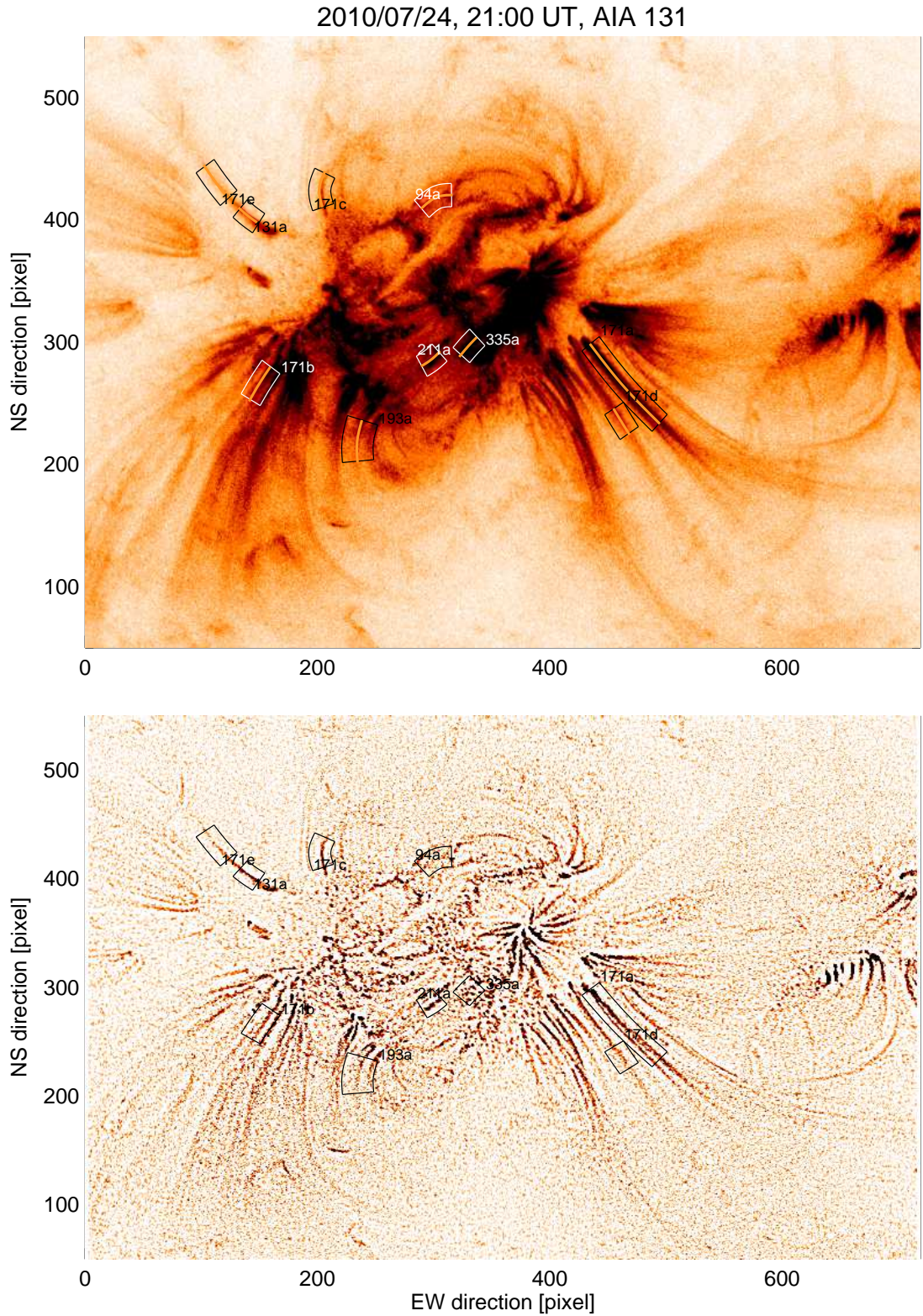}}
\caption{Partial AIA image observed at 171 \ang\ on 2010 July 24, 21:00 UT,
rendered with inverse greyscale (top) and as a highpass-filtered version
(by subtracting the original image smoothed with a boxcar of $7\times 7$
pixels). The location of the extracted loop segment boxes are indicated,
which are curved arrays aligned with the loop spines at the 10 selected
loop locations. The background fluxes are interpolated between the boundaries
of the boxes on both sides of the loop spines.}
\end{figure}

\begin{figure}
\centerline{\includegraphics[width=\textwidth]{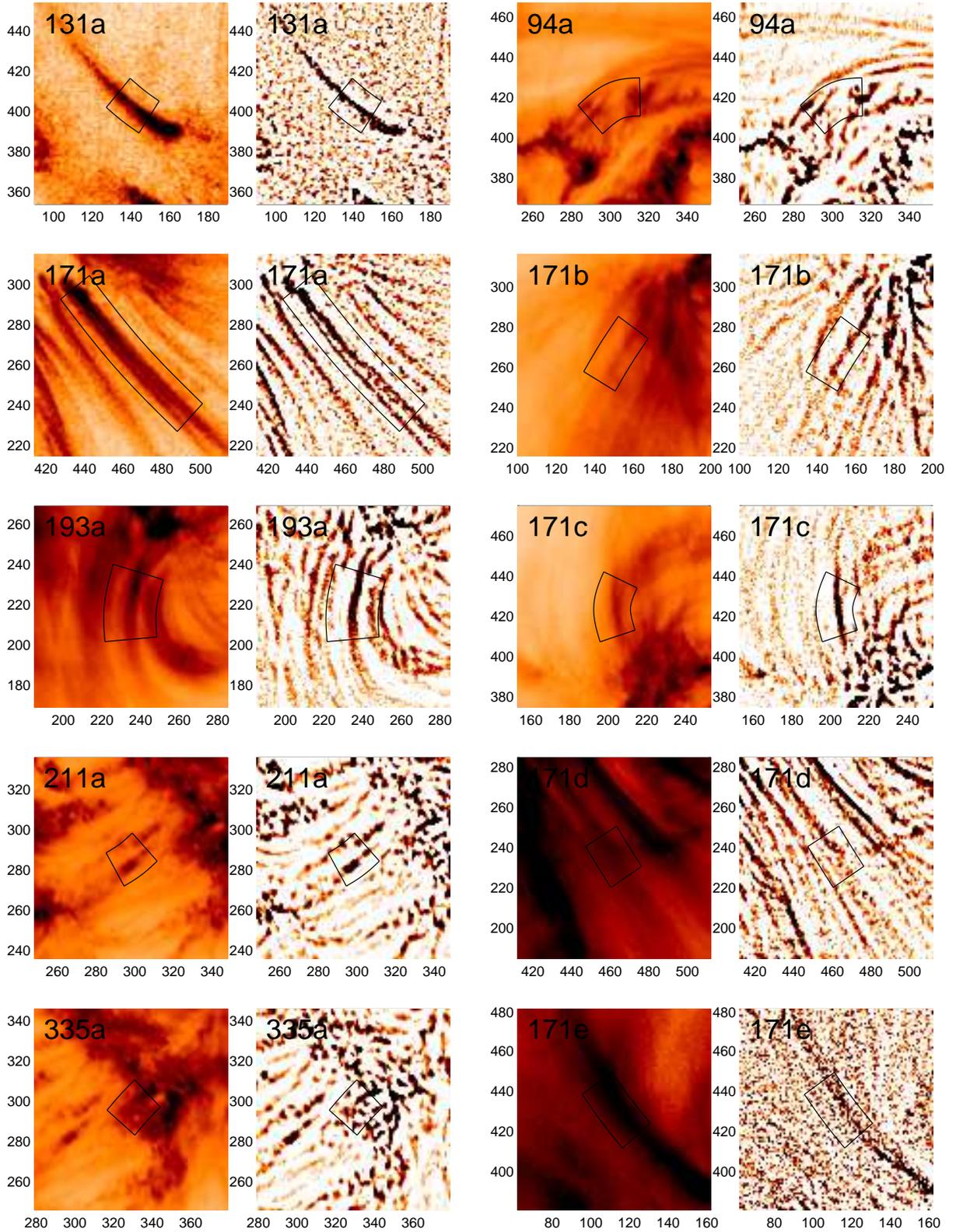}}
\caption{Enlargements of $100 \times 100$ pixel subimages around the
selected 10 loops, extracted from the same AIA 171 \ang\ image as shown
in Fig.~1, observed on 2010 July 24, near 21:00 UT.}
\end{figure}

\begin{figure}
\centerline{\includegraphics[width=\textwidth]{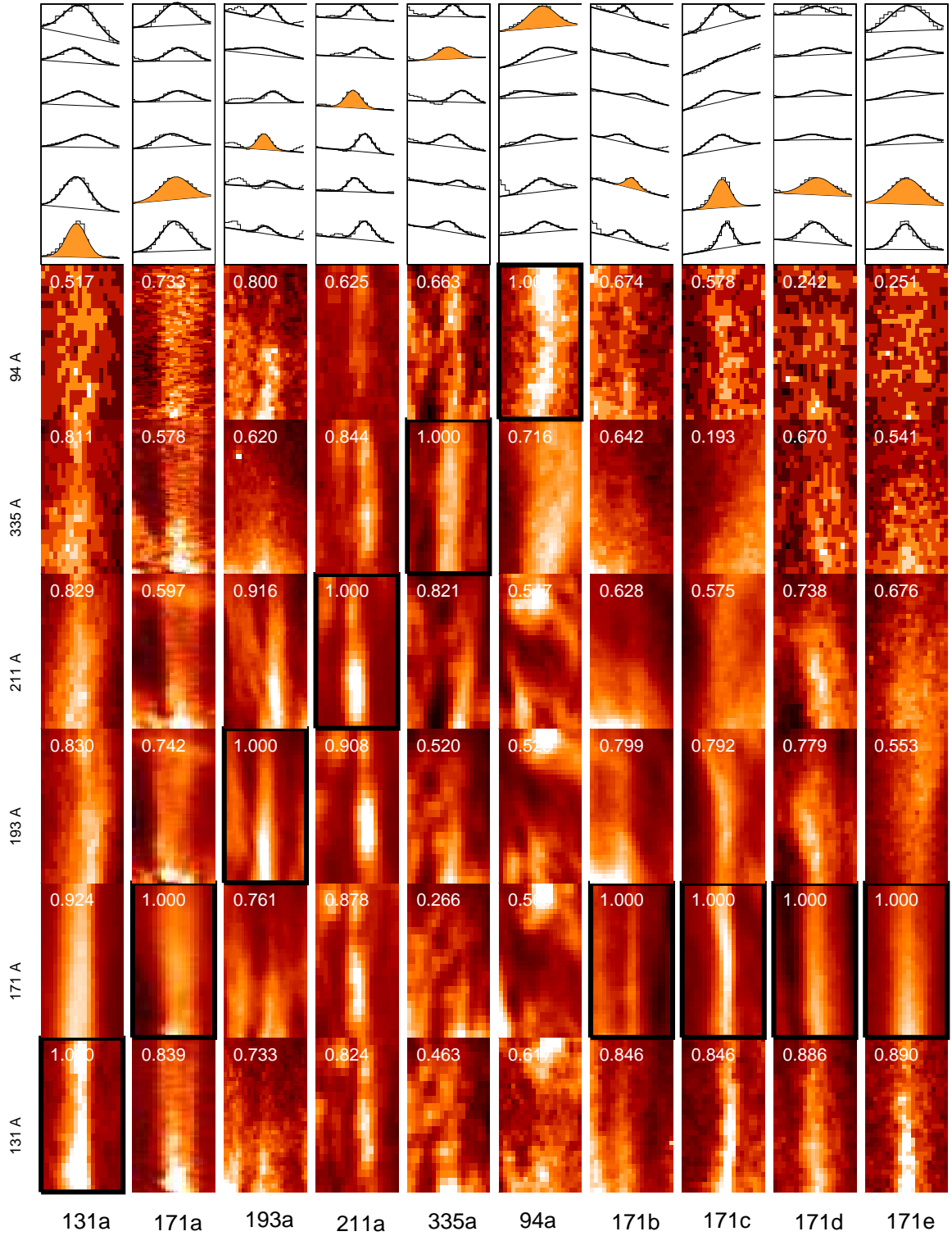}}
\caption{Stretched-out loop segments aligned with the loop spine (in vertical
direction) for 10 loop segments at the first of the 10 time intervals, shown
for all 6 coronal filters. The cross-sectional flux profiles $F_{\lambda}(x)$ 
in all 6 filters are shown in the top part. The wavelengths in which each loop 
was detected are indicated with a black frame and with a grey-colored cross-section.
The cross-correlation coefficients between the partial image of each wavelength
with the wavelength of detection are indicated in each partial image.}
\end{figure}

\begin{figure}
\centerline{\includegraphics[width=\textwidth]{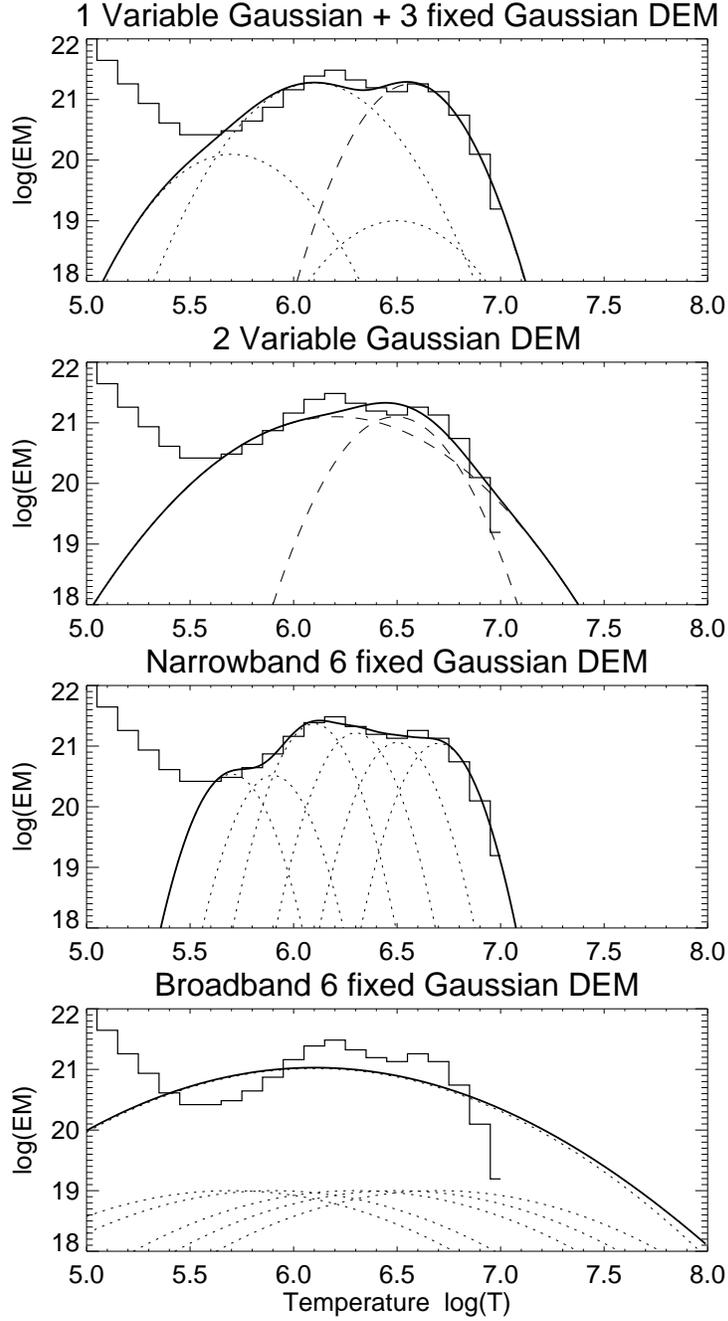}}
\caption{An observed DEM (histogram) from Brosius et al.~(1996), labeled as AR 93,
is fitted with our four different DEM parameterizations (Eq.~7).
The best-fit DEM models are marked with thick solid lines, the
variable Gaussian components (with three free parameters 
($EM_i, T_i, \sigma_{T}$) are marked with dashed curves, and the
fixed Gaussians ($EM_i$ variable, but $T_i$ and $\sigma_{Ti}$ are fixed)
are marked with dotted curves. Note that the first 3 models with narrowvand
Gaussians ($\sigma_T \ge 0.1$) can represent the observed DEM with reasonable 
accuracy, while the broadband Gaussian model $\sigma_T=0.5$ fails (bottom).}
\end{figure}

\begin{figure}
\centerline{\includegraphics[width=\textwidth]{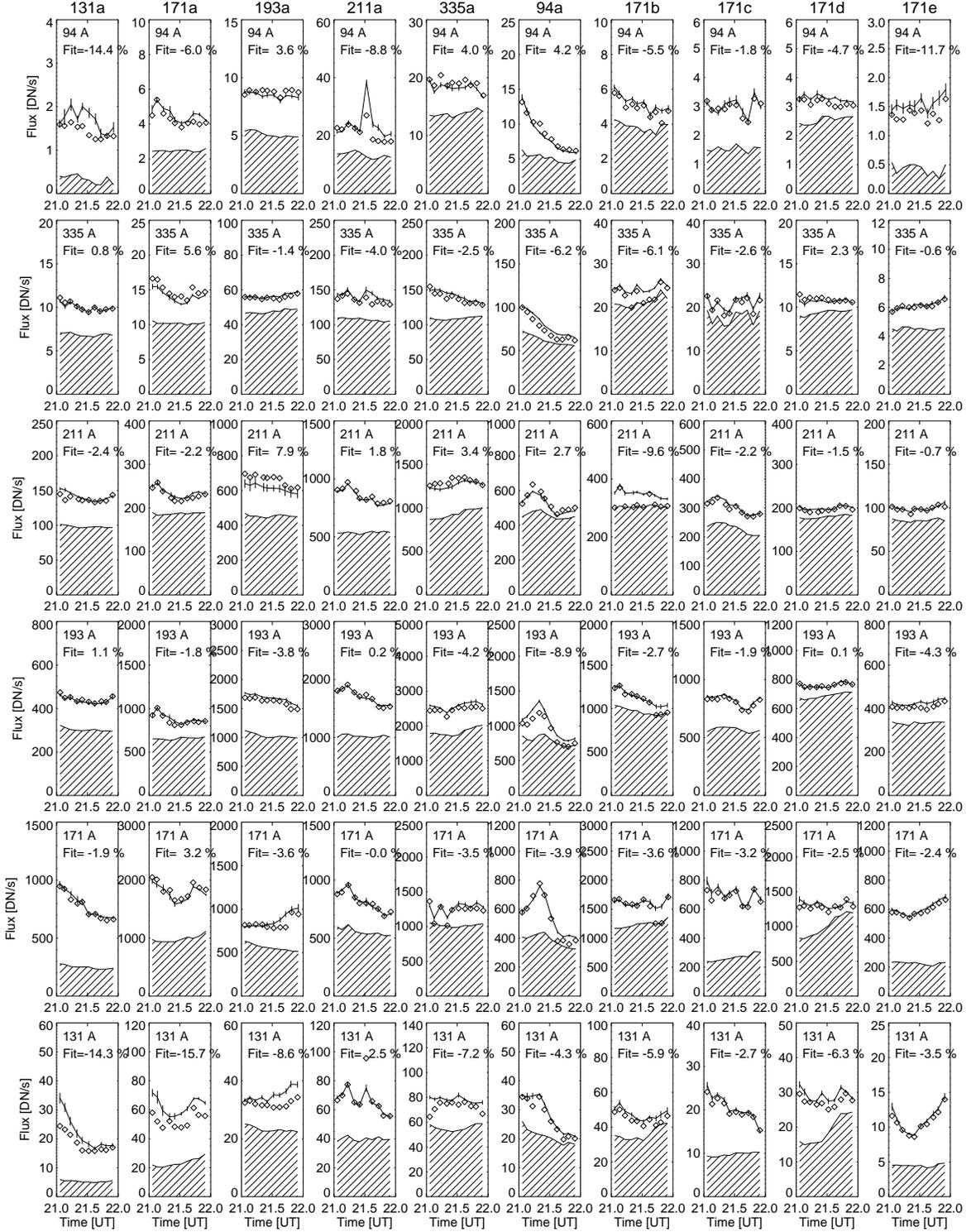}}
\caption{Time profiles of background fluxes $B_{\lambda}(t)$ (hatched),
total fluxes $F_{\lambda}(t)$ (solid curve with error bars), and best fits
$F^{mod}_{\lambda}(t)$ with the 2-Gaussian DEM model 
in 6 wavelengths (sequence in vertical direction) 
for 10 different loop positions (sequence in horizontal direction). The wavelength
$\lambda$ and fit accuracy (in percents) is indicated in each panel.}
\end{figure}

\begin{figure}
\centerline{\includegraphics[width=\textwidth]{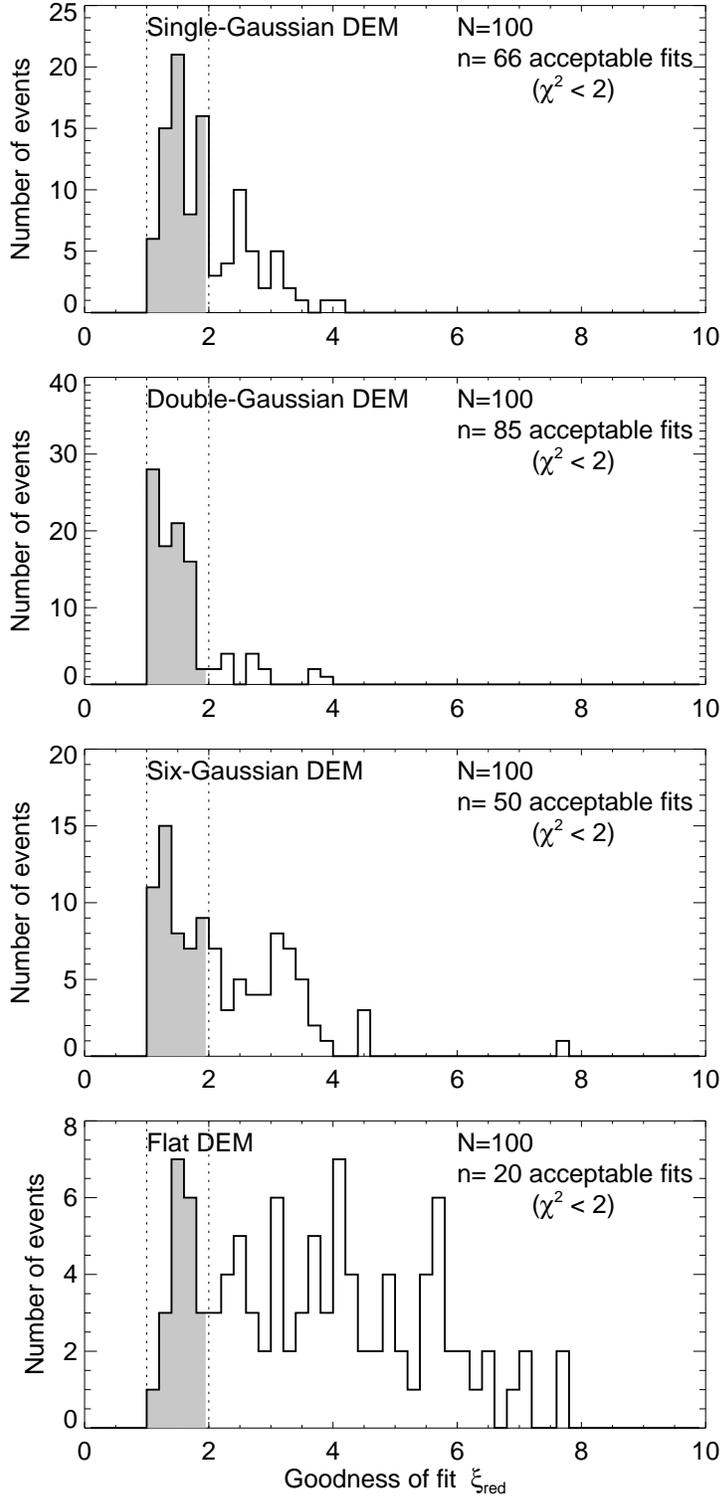}}
\caption{Histograms of the goodness-of-fit $\chi^2$-criterion for the 4 different
DEM models: single-Gaussian (top); double-Gaussian (second); six-Gaussian (third);
and flat DEM (bottom). Note that all models yield $\chi^2 > 1$, but only the
double-Gaussian model contains most values (81\%) within an acceptable 
range of $\chi^2 < 2$.}
\end{figure}

\begin{figure}
\centerline{\includegraphics[width=\textwidth]{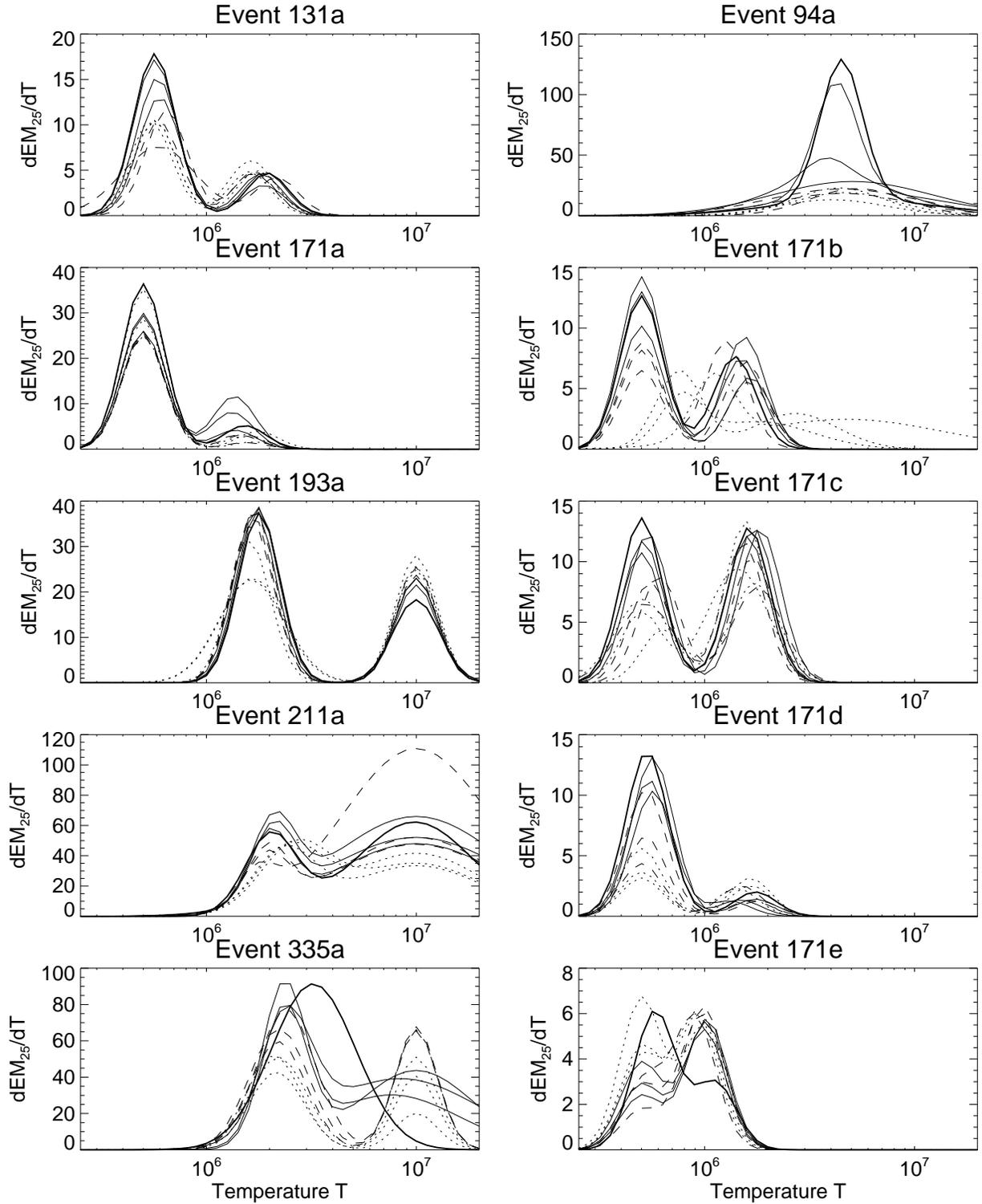}}
\caption{Differential emission measure (DEM) distributions fitted with the
2-Gaussian DEM model for the 10 different locations (each frame) in time steps
of 6 minutes during the time interval of 2010 Jul 24, 21:00-22:00 UT. The first
time frame corresponds to the detection time of a loop (first time step),
time steps 2-4 are shown with solid linestyle, time steps 5-7 with dashed linestyle,
and time steps 8-10 with dotted linestyle. Note the continuous evolution of the
best-fitt DEMs.}
\end{figure}

\begin{figure}
\centerline{\includegraphics[width=\textwidth]{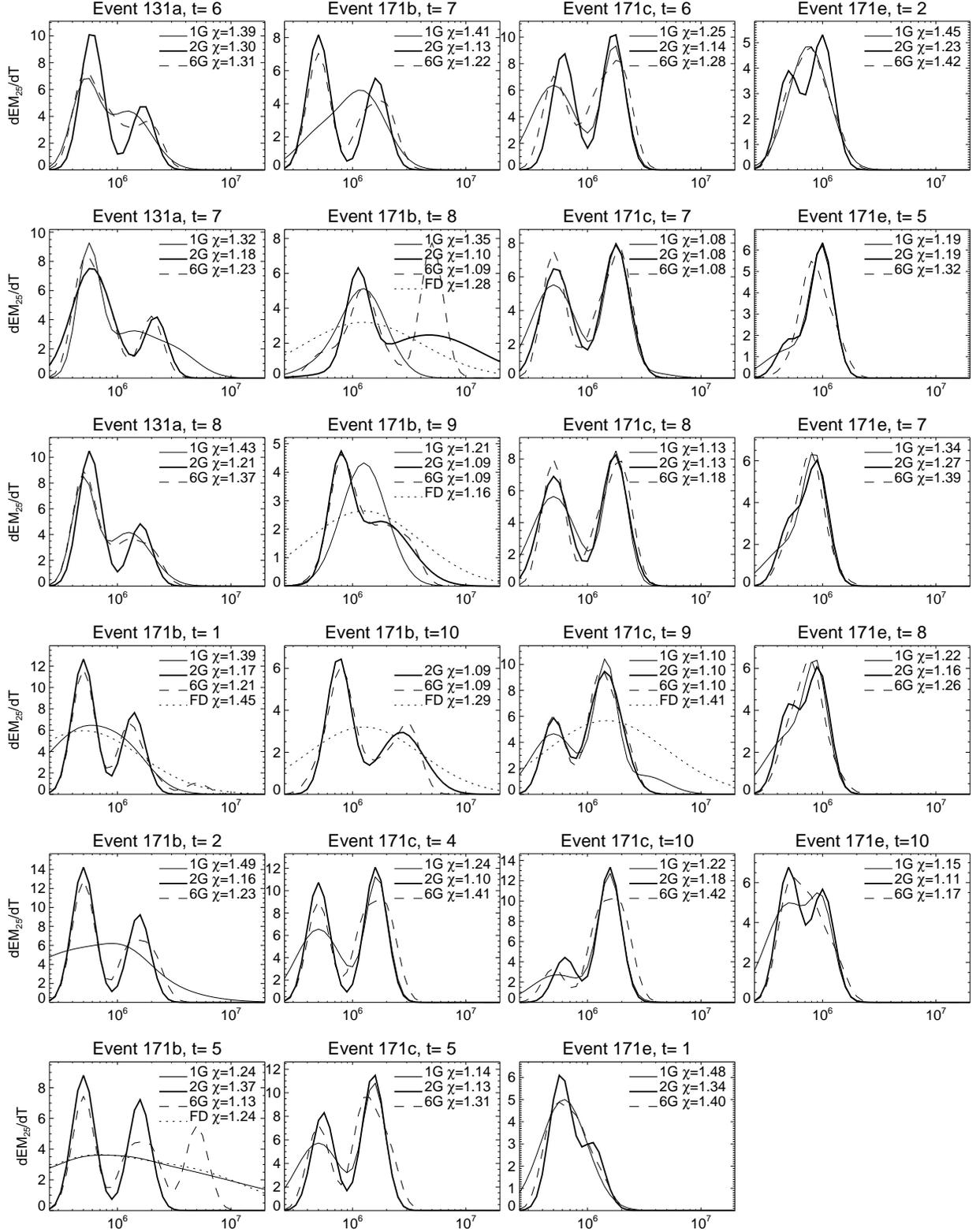}}
\caption{DEM solutions are shown for events where at least 3 different
DEM models yield a goodness-of-fit of $\chi^2 \le 1.5$, which is the
case for 23 out of the 100 events. The 4 different DEM models are: 
single-Gaussian (thin solid line), double-Gaussian (thick solid line), 
narrowband six-Gaussian (dashed line), and broadband six-Gaussian DEM model
(dotted line).}
\end{figure}

\begin{figure}
\centerline{\includegraphics[width=\textwidth]{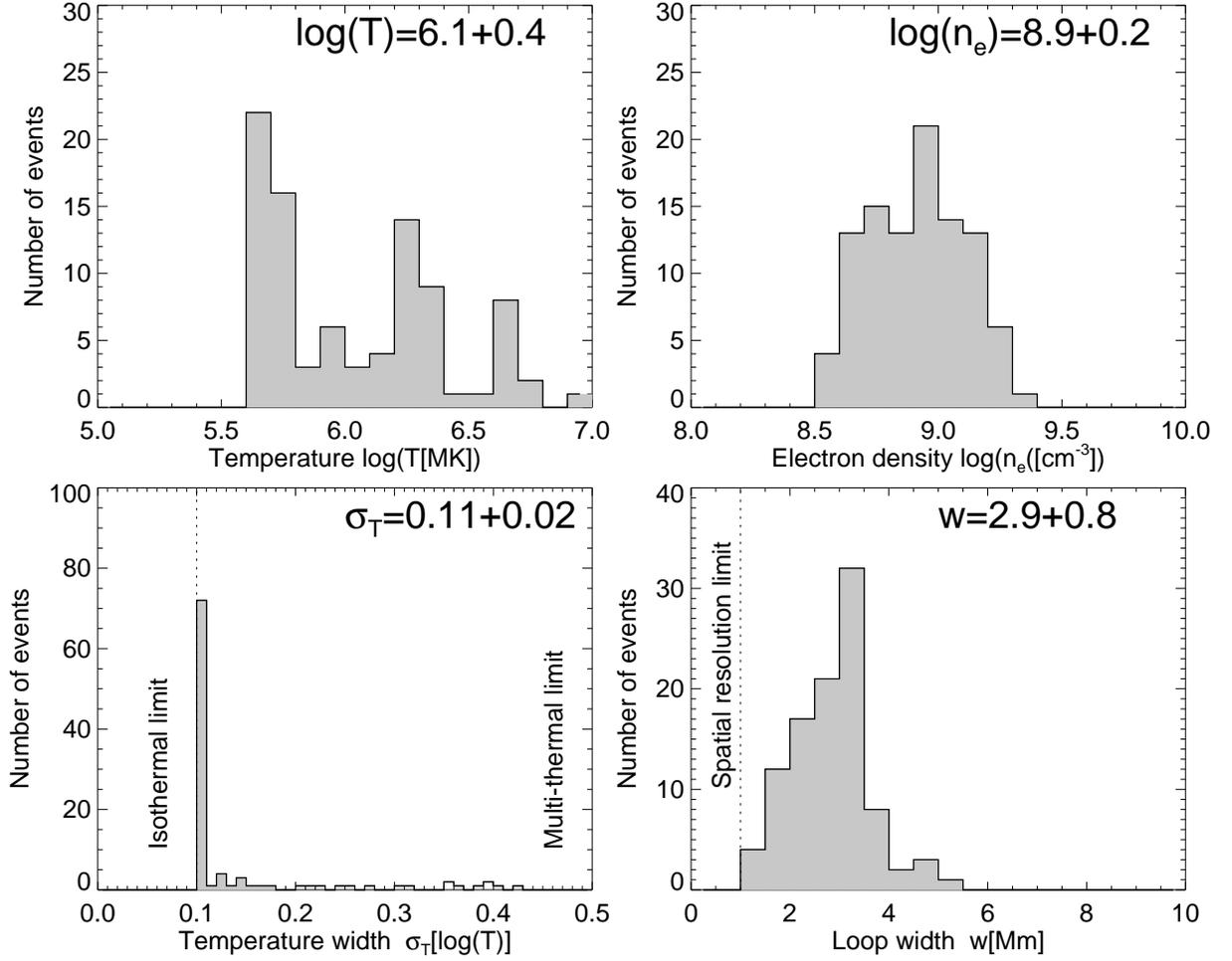}}
\caption{Parameter distributions of the principal DEM component measured with the
2D-Gaussian DEM model: the loop temperature $log(T)$ (top left), the electron
density $log(n_e)$ (top right), the logarithmic temperature width $\sigma_T$ 
(bottom left), and the loop width $w$ (bottom right).}
\end{figure}

\begin{figure}
\centerline{\includegraphics[width=\textwidth]{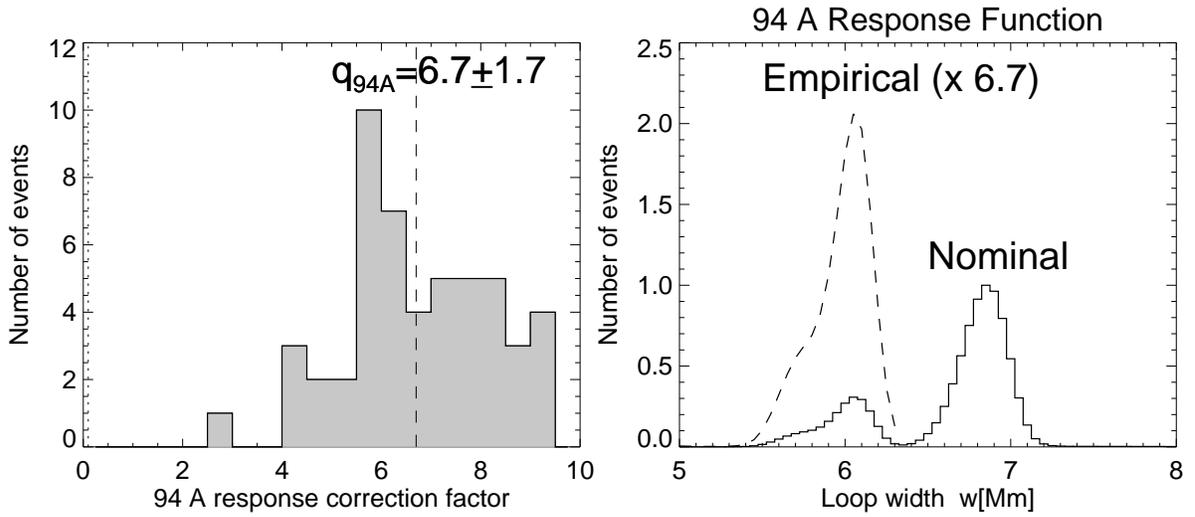}}
\caption{Empirical correction of AIA94 \ang\ response function for the
low-temperature response at $log(T) < 6.3$ (right panel). 
The low-temperature response
needs to be boosted by a factor of $6.7\pm1.7$ according to best-fit DEM
solutions obtained from 100 DEM fits using a 2-Gaussian DEM model
(left panel).}
\end{figure}

\begin{figure}
\centerline{\includegraphics[width=\textwidth]{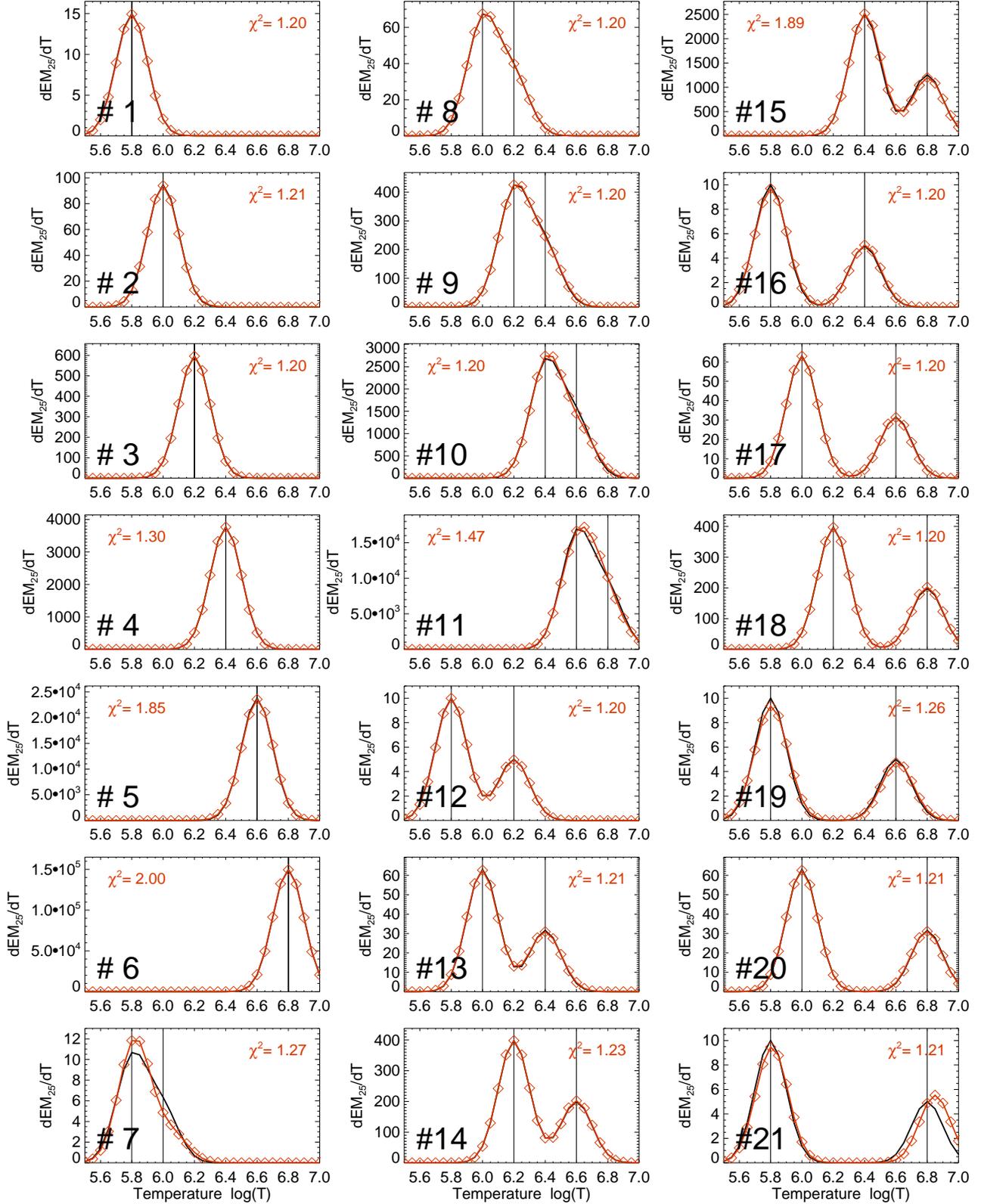}}
\caption{A set of 21 simulated DEM distributions with one narrow (\#1-\#6) 
or two Gaussian peaks (\#7-\#21), shown as black curves with the location
of the temperature peaks marked with vertical lines. These DEMs have been
convolved with the instrumental response functions of the 6 AIA filters
for $n_w=20$ pixels of a Gaussian loop cross-section, with photon noise 
added, and inverted with forward-fitting of a double-Gaussian DEM model 
(red curves and diamonds). The $\chi^2$-squares indicate the goodness 
of the best fit (see Appendix A).}
\end{figure}

\end{document}